\documentclass[default,referee]{sn-jnl}

\usepackage{graphicx}%
\usepackage{multirow}%
\usepackage{amsmath,amssymb,amsfonts}%
\usepackage{amsthm}%
\usepackage{mathrsfs}%
\usepackage[title]{appendix}%
\usepackage{xcolor}%
\usepackage{textcomp}%
\usepackage{manyfoot}%
\usepackage{booktabs}%
\usepackage{algorithm}%
\usepackage{algorithmicx}%
\usepackage{algpseudocode}%
\usepackage{listings}%

\usepackage{tikz,pgf}
\usepackage{float}
\usepackage{caption}
\usepackage[english]{babel}
\usepackage{enumerate}
\usepackage{graphicx,epsfig,color}
\usepackage{listings}
\usepackage{lscape}
\usepackage{mathrsfs}
\usepackage{multirow,array}
\usepackage{tabularx}
\usepackage{wrapfig}
\usepackage{eurosym}
\theoremstyle{plain}
\theoremstyle{definition}
\newtheorem{lemma}{Lemma}
\newtheorem{assumption}{Assumption}

\newcommand{\beq}{\begin{equation}}
\newcommand{\eeq}{\end{equation}}

\newcommand{\lf}{\left}
\newcommand{\rg}{\right}

\def\min{\operatorname{min}}
\def\max{\operatorname{max}}

\newcommand{\R}{\mathbb R}
\newcommand{\E}{\mathbb E}
\newcommand{\PP}{\mathbb P}

\theoremstyle{thmstyleone}%
\newtheorem{theorem}{Theorem}
\newtheorem{proposition}[theorem]{Proposition}%

\theoremstyle{thmstyletwo}%
\newtheorem{example}{Example}%
\newtheorem{remark}{Remark}%

\theoremstyle{thmstylethree}%

\begin{document}

\title[A Profit-Maximizing Strategy for Advertising on the e-Commerce Platforms]{A Profit-Maximizing Strategy for Advertising on the e-Commerce Platforms}


\author[1]{\fnm{Lianghai} \sur{Xiao}}

\author*[2]{\fnm{Yixing} \sur{Zhao}}\email{yzhao@gdufs.edu.cn}

\author[3]{\fnm{Jiwei} \sur{Chen}}

\affil[1]{\orgdiv{School of Mathematics}, \orgname{South China University of Technology}, \orgaddress{\city{Guangzhou},  \country{China}}}

\affil*[2]{\orgdiv{School of Finance}, \orgname{Guangdong University of Foreign Studies}, \orgaddress{\city{Guangzhou}, \postcode{510006},  \country{China}}}

\affil[3]{\orgdiv{Data Analysis Department}, \orgname{ShenZhen TingZhu Health Management Co.,Ltd.}, \orgaddress{\city{ShenZhen}, \country{China}}}


\abstract{The online advertising management platform has become increasingly popular among e-commerce vendors/advertisers, offering a streamlined approach to reach target customers.  Despite its advantages,  configuring advertising strategies correctly remains a challenge for online vendors, particularly those with limited resources. Ineffective strategies often result in a surge of unproductive ``just looking'' clicks, leading to disproportionately high advertising expenses  comparing to the growth of sales. In this paper, we present a novel profit-maximing strategy for targeting options of online advertising. The proposed model aims to find the optimal set of features to maximize the probability of converting targeted audiences into actual buyers. We address the optimization challenge by reformulating it as a multiple-choice knapsack problem (MCKP). We conduct an empirical study featuring real-world data from Tmall to show that our proposed method can effectively optimize the advertising strategy with budgetary constraints.}




\maketitle

\section{Introduction}\label{sec:intr}

In the past decade, the online shopping market has witnessed remarkable growth, leading to the rise of online advertising \citep{goldfarb2014different}. Traditionally, advertisers negotiated long-term contracts with publishers to purchase online display advertising spots. In recent years, advertisers now have access to real-time feedback on their online advertising endeavors, allowing them to precisely target specific audiences \citep{evans2008economics, goldfarb2011online}.

There has been a surge in the adoption of advertising manager in the e-commerce platforms. Examples of these platforms include Google Ads, Microsoft Advertising, Meta's Ads Manager, and Alibaba's DMP. These platforms are designed to streamline the process of running digital advertising campaigns, making it easier for advertisers to reach their targeted audience and achieve their marketing goals. They offer a wide array of online advertising opportunities on their own social networks and online shopping platforms. For instance, Meta's Ads Manager empowers businesses to create and oversee ad campaigns across Facebook, Instagram, and various other social media of Meta network. Similarly, Alibaba's DMP facilitates e-commerce sellers to advertise on a vast consumer base within Tmall and Taobao.

Upon customer registration on a webpage, adhering to the privacy policy agreement, their information is collected alongside their behavioral data. 
Subsequently, customers are categorized into various dimensions such as demographics, interests, and location. This accumulated information, in combination with the target options set by advertisers, facilitates the precise delivery of ads to the intended audience. By leveraging this data-driven approach, advertisers can continually monitor the results of their advertising strategies and make necessary adjustments directly on the platform, ensuring their campaigns remain timely and effective.

Nevertheless, configuring an efficient advertising strategy using these comprehensive all-in-one tools can be a significant challenge for small or medium-sized enterprises (SMEs). Many online advertising platforms adopt charging formulas such as cost-per-click (CPC) or cost-per-thousand-impressions (CPM)
\footnote{CPC and CPM are two common pricing models used in online advertising. 
CPM is a pricing model where advertisers pay for every 1,000 times their ad is displayed to a user, regardless of whether the user clicks on the ad or not. CPM is often used for brand awareness campaigns where the goal is to get the ad seen by as many people as possible, rather than generating clicks.}. 
An inadequate advertising strategy could result in a calamity, particularly if the CPC or CPM ads attract a surge of ``just looking'' customers. As an advertiser's ultimate objective is to generate revenue and attract new customers, which can only be achieved through compelling and effective ad strategies that drive conversions.

In this paper, we present a novel model for elucidating  an advertiser's advertising strategy. In our model, customers are characterized by various features and attributes. Our  goal is to identify the optimal set of features that maximizes the conversion rate of customers with the chosen features. Under some mild assumptions, the model is ultimately reformulated to a multiple choice knapsack problem for which numerous algorithms are available to find the optimal solution. Our study provides a practical procedure for the advertisers to identify their optimal advertising strategy. The attainability of the optimal strategy is only contingent upon knowing the percentage of  customers with specific features.

Online advertising has been extensively studied in recent literature from various perspectives, including revenue optimization \cite[see, e.g.,][]{balseiro2014yield,balseiro2015repeated}, algorithms \cite[see, e.g.,][]{miralles2018novel}, and bidding strategies \cite[see, e.g.,][]{lee2013real, grigas2017profit, lobos2018optimal, liu2020effective}.  \citet{grigas2017profit} and \citet{lobos2018optimal} developed optimization models for managing demand-side platforms (DSPs) with the goal of maximizing the platform's profit by choosing the optimal real-time bidding (RTB) strategy given a budget constraint. The strategy involves impression type allocation and bidding prices. \citet{lee2013real} presented an RTB strategy that smooths the budget allocation over time and maximizes conversion performance. Our research differs from these papers in two aspects. Firstly, we aim to maximize the revenue of advertisers rather than platforms or publishers. Secondly, we do not incorporate price bidding in our model as advertisers usually deliver their ads to targeted audiences by directly purchasing display slots at pre-set prices based on their specific targeting criteria. This approach is particularly relevant in highly competitive markets and focus on empowering SMEs with optimal strategies.

Our work is also related to click-through-rate (CTR) estimation for advertising campaigns. Many researchers \cite[e.g.,][]{miralles2018novel, liu2020effective} aim to optimize advertising campaigns with the aid of CTR estimation. CTR estimation plays a crucial role in optimizing advertising campaigns, and various prediction and estimation methods have been well-studied. Logistic regression or generalized linear models have been widely used for CTR prediction based on customer features \cite[see, e.g.,][]{richardson2007predicting, lee2018estimating} owing to their easy implementation. Furthermore, more advanced techniques have been developed, such as tensor factorization models \citep{shan2014ctr}, field-aware factorization machine \citep{juan2016field}, Product-based Neural Network \citep{qu2016product}, and factorization machine supported neural network \citep{zhang2016deep}. These approaches focus on predicting the probability of a customer clicking on an ad given their features, and the optimization models revolve around CTR. In some senses, our model is different from these methods but is similar to investment models that optimize return on investment (ROI).
    
The contribution of this paper is threefold. 
Firstly, so far as we know, this is the first study to specifically address the issue of optimal advertising strategy for online advertisers that is only a few accessible information points from online advertising management platform. Prior research has focused on real-time bidding strategies for platforms or relied on challenging CTR prediction methods. These approaches may pose challenges for advertisers to obtain relevant data and are less practical. 
Secondly, we propose a procedure that offers a practical guideline for advertisers, particularly SMEs, to identify their optimal advertising strategies. The method is meticulously described  and can be readily implemented. 
Thirdly, the study conducts a numerical experiment using real-world data obtained from an enterprise, serving as empirical evidence of the efficacy of the proposed model. This validation showcases the effectiveness of the approach in real-world settings.

The subsequent sections of this paper are organized as follows. The advertising management platform is described in detail in Section \ref{sec: plat}. Section \ref{sec:mod} introduces a comprehensive model that describes the advertising strategy for an e-commerce seller.  In Section \ref{sec:remod}, the model is further reformulated to a solvable integer programming problem, provided that some mild assumptions are included.  Section \ref{sec:exp} presents a step-by-step procedure to numerically determine the optimal advertising strategy. An empirical study featuring real-world data from an SME on Tmall is provided to illustrate the effectiveness of the proposed model. Finally, the paper concludes with closing remarks in Section \ref{sec:conc}.

\section{Advertising management platforms}\label{sec: plat}
An advertising management platform is a platform that aids advertisers in managing their campaigns across many channels. 
Key features of an advertising management platform may include:
\begin{itemize}
\item \emph{Ad Creation}: The platform enables advertisers to create and design their ads with various formats, including text, images, videos, and interactive media.

\item \emph{Targeting Options}: Advertisers can define specific criteria, such as demographics, interests, behavior, and location, to target their ads to a relevant audience.

\item \emph{Campaign Management}: advertising managements allow users to set budgets, select bidding strategies, and schedule ad campaigns to run at specific times.

\item \emph{Performance Tracking}: The platform provides real-time analytics and reports on the performance of ad campaigns, including impressions, clicks, conversions, and return on investment (ROI).
\end{itemize}

We aim to determine effective advertising strategies for advertisers, leveraging the \emph{targeting options} offered by an advertising management platform. Specifically, the platform provides advertisers with a wide range of criteria to define their targeted audience, which can be classified into: 
\begin{itemize}
\item \emph{Consumption Behavior} - This category measures the shopping activities of customers on the platform. The features in this category may include consumption frequency, monthly expenditure, purchasing power level, etc.

\item \emph{Interests} - This category reflects the interests of customers on the platform, based on the products they have liked or favorited.

\item \emph{Demographic Features} - This category includes customer demographics, locations, and other information provided by customers or estimated by the platform.

\item \emph{Behavioral Preferences} - This category encompasses customer behavior preferences, such as browsing patterns, frequently used devices, shopping preferences, and more.
\end{itemize}

\begin{figure}[h]
\centering 
\begin{tikzpicture} [scale=0.6, every node/.style = {shape=rectangle, rounded corners,
    draw, align=center,
    top color=white, bottom color=blue!20}]
\node[shape=rectangle, draw=black, minimum width=3cm, minimum height=1.5cm] (1) at (-6,0) {Create a new campaign};
\node[shape=rectangle, draw=black, minimum width=3cm, minimum height=1.5cm] (2) at (0,3) {Choose targeted audience};
\node[shape=rectangle, draw=black, minimum width=3cm, minimum height=1.5cm] (3) at (8,3) {Run ads};
\node[shape=rectangle, draw=black, minimum width=3cm, minimum height=1.5cm] (4) at (8,-3) {Get reports};
\node[shape=rectangle, draw=black, minimum width=3cm, minimum height=1.5cm] (5) at (0,-3) {Adjust campaign};

\path[-latex,line width=3pt] (1.north) edge (2.west);
\path[-latex,line width=3pt] (2.east) edge (3.west);
\path[-latex,line width=3pt] (3.south) edge (4.north);
\path[-latex,line width=3pt] (4.west) edge (5.east);
\path[-latex,line width=3pt] (5.north) edge (2.south);

\end{tikzpicture}
\caption{The procedure of  using an advertising management platform}
\label{Fig:procedure}
\end{figure}
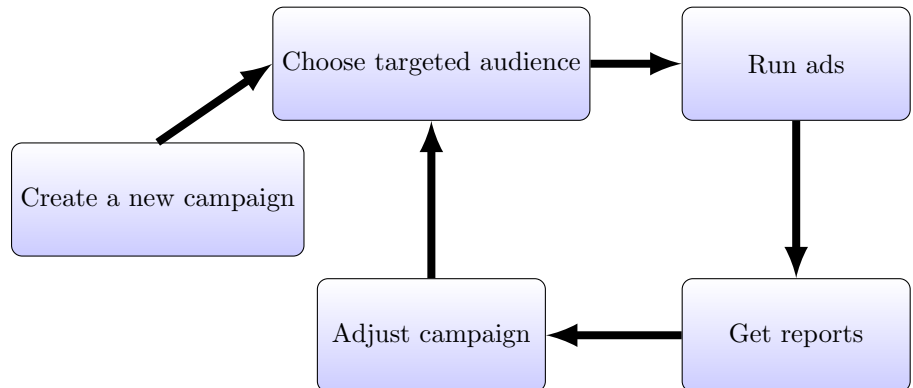

We present Figure \ref{Fig:procedure} to visually depict the process of an online advertiser using these platforms. In this paper, we refer to the advertiser's choices of criteria as the \textit{advertising strategy}. 

\begin{figure}[h]
  \centering
  \includegraphics[scale=0.3]{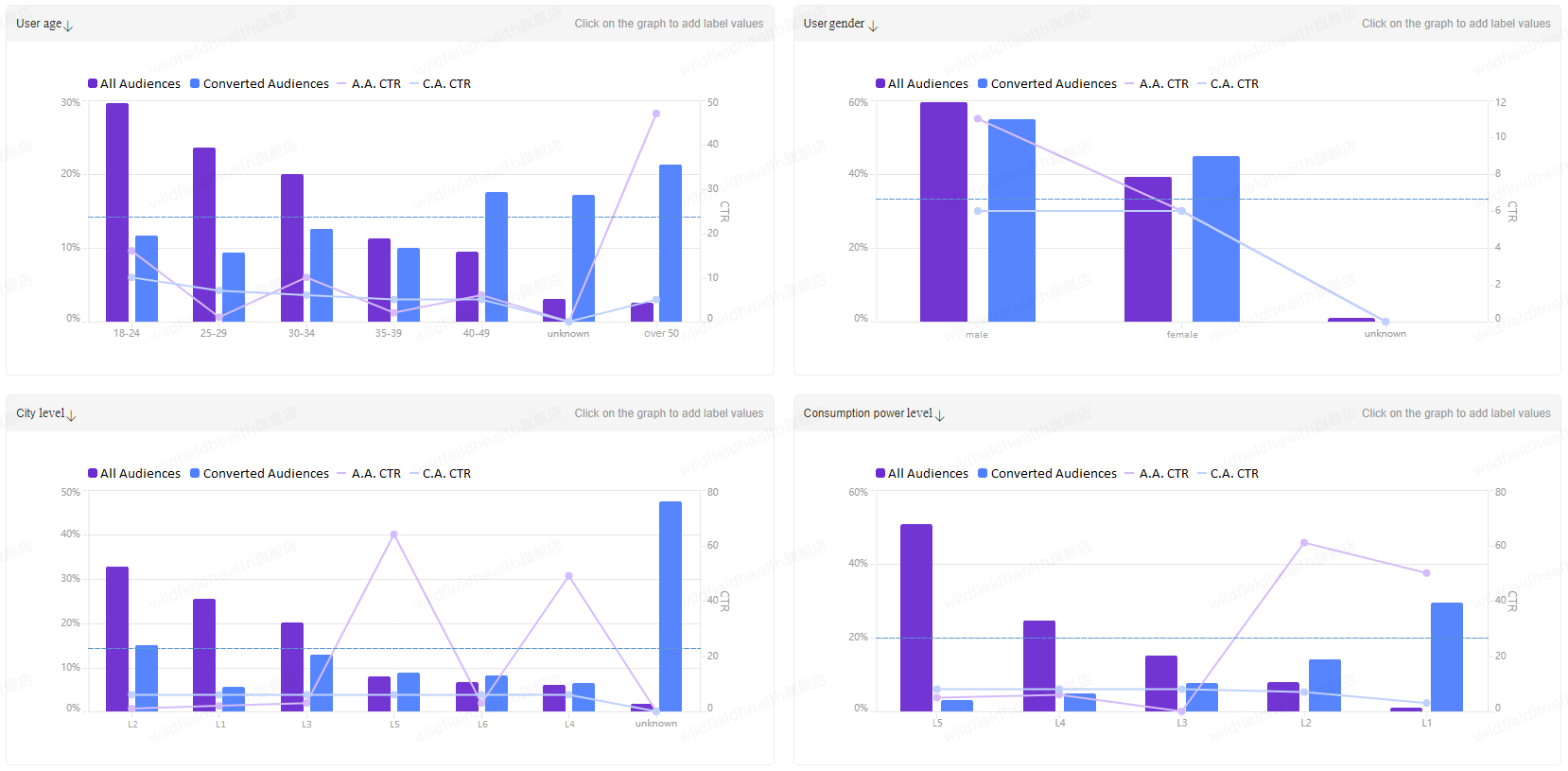}
  \caption{\small Screenshot of the interface of Tmall's ads manager tool}
  \label{fig:ss}
  {\footnotesize Note: This figure shows an interface of Tmall's advertising management platform for vendors. Such analysis panel can be found in many e-commercial platforms such as Tmall, Amazon, TikTok, etc. The panel depicts the statistical data of customers' age, gender, city level, and consumption power level. The purple bars illustrate the distributions of types of these features for the real customers (corresponding to $P_{i,k}$ in \eqref{eq:pik}), and the blue bars are for the potential customers (corresponding to $Q_{i,k}$ in \eqref{eq:qik}).\par} 
\end{figure}

To determine the potential customers they can attract, advertisers often conduct \emph{trial-run} advertisements on e-commerce platforms. Typically, these platforms require a certain level of advertising expenditure before granting access to their advertising management platforms. During the trial-run period, the product is displayed to audiences of all types, with the reach depending on the advertiser's budget. Gathering feedback data from audiences who become actual customers is a critical aspect of this process. The feedback data is then used to collect and analyze feature data, allowing the platform to generate statistical insights into customer preferences. A larger customer exposure during the trial run generally leads to less biased statistical results, as it allows the product to be seen by a more diverse audience. advertising management platform can provide a panel with visualized data to advertisers (see figure \ref{fig:ss}).
These data are updated regularly, allowing advertisers to dynamically tailor their advertising strategy to effectively reach their target audience. In this paper, we will demonstrate the process of utilizing the data available on the advertising management platform to construct an optimal advertising strategy for advertisers.

\section{Model Fundations}\label{sec:mod}
We consider the case that a profit-maximizing advertiser
involved in the online sale of a single product. The product's price is denoted as $p$, and its production cost is represented by $c$. The quantity of product sold, denoted as $q$, is influenced by two key factors: the ad campaign expenditure budget, denoted as $e$, and the chosen advertising strategy, denoted as ${\cal S}$. 
Note that we assume $p>c\ge 0$ and $q\ge 0$, ensuring that the product's selling price is higher than its production cost, and non-negative quantities are sold. Additionally, we assume that the maximum possible sales volume will not exceed the scale-free capabilities, such that the cost per product $c$ remains constant regardless of the production volume. Moreover, the price $p$ is predetermined, leaving the advertiser with the task of optimizing both $e$ and ${\cal S}$ to maximize its profit:
\begin{equation}\label{eq:obj0}
	\underset{ 	e,{\cal S}}{\max}~ f(e;{\cal S}):=q(e;{\cal S})(p - c) - e.
\end{equation} 

A plausible approach for optimization is to employ an alternative minimization process, where we iteratively optimize the variables ${\cal S}$ and $e$ in turns. Alternatively, the advertiser can design its advertising strategy ${\cal S}$ with a pre-determined budget. Suppose that the campaign expenditure budget is fixed to $e_0$, the ${\cal S}$-subproblem is formulated as follows:
\begin{align}\label{eq:optb}
{\cal S}_0^* & := \underset{{\cal S}}{\arg\max}~ q(e_0;{\cal S})(p - c) - e_0 \\\nonumber
	& =\underset{{\cal S}}{\arg\max}~ q(e_0;{\cal S}),
\end{align}
then the profit-maximizing problem \eqref{eq:obj0} is equivalent to maximizing the sales $q$ with a given expenditure budget $e_0$.  

\textbf{Optimizing the advertising strategy.} 
As empowered by the advertising management platforms, advertiser has the flexibility to target specific audiences based on various features. Consequently, the advertising strategy of an advertiser ${\cal S}$ can be characterized by a sequence of features relating to audiences. We denote $ {\cal F}:=\{F_1,F_2,\dots,~F_n\}$ the set of $n$ numerical and/or categorical features available on the advertising management platform, examples of which include \texttt{gender, age, location, operating system, browser}, etc. Within each feature $F_i$, there exist $m_i$ distinct types, denoted by $F_i= \{t_{i,1},~t_{i,2},\dots,~t_{i,m_i}\}$ for $i\in[n]:=\{1,2,\dots,n\}$. The number of elements in $F_i$ is represented by  $|F_i|=m_{i}$. For instance, \texttt{age=\{male, female, unknown}\} has $|$\texttt{age}$|=3$. Note that when audiences choose not to disclose their personal information, due to the privacy policies, they are considered as part of the \texttt{unknown} category.
For the convenience of notation, we define the following event:
\[
	(Buy) = (Customer~will~buy~the~product~after~seeing~the~ad).
\]

\begin{figure}[h]
\centering 
\begin{tikzpicture}[scale=0.8, grow=right, level distance=12em,
  level 1/.style={sibling distance=8em},
  level 2/.style={sibling distance=2em},
  every node/.style = {shape=rectangle, rounded corners, minimum width=7em, scale=0.8,
    draw, align=center,
    top color=white, bottom color=blue!20}]
  \node {$\cal F$}
    child { node {$F_{n}$} 
       child { node {$t_{n,m_{n}}$} }
       child { node {...} }
       child { node {$t_{n,2}$} }
       child { node {$t_{n,1}$} }}
    child { node {...} }
    child { node {$F_{2}$: age} 
      child { node {$t_{2,5}$: unknown} }
      child { node {$t_{2,4}$: $>40$} }
      child { node {$t_{2,3}$: $31\sim40$} }
      child { node {$t_{2,2}$: $19\sim 30$} }
      child { node {$t_{2,1}$: $\leq 18$} }}
    child { node {$F_{1}$: gender} 
       child { node {$t_{1,3}$: unknown} }
       child { node {$t_{1,2}$: female} }
       child { node {$t_{1,1}$: male} }};
\end{tikzpicture}
\caption{The flow chart for the relationships about $\cal F$, $F_{i}$ and $t_{i,k}$}
\label{Fig:flowchart}
\end{figure}
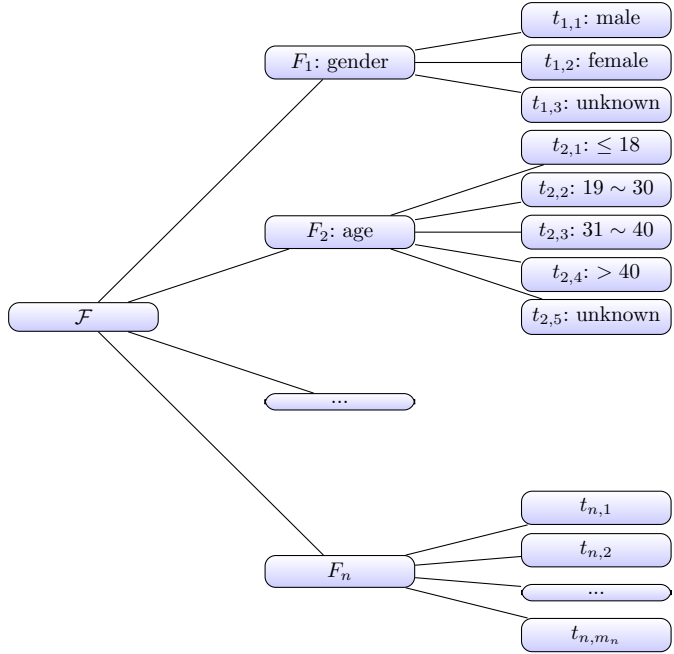

The statistical data on audiences' features are subject to periodic updates, as the group of active users  on an e-commerce platform is dynamic. Moreover, trial runs may result in a certain level of bias in the statistical results. Therefore, it is necessary for an advertiser to regularly update its optimal advertising strategies to align with these changes. However, since our focus is on finding the optimal advertising strategy, we will not delve into the technical details of these updates.

\textbf{Explanatory variables.} 
Following the trial-run advertisement, the advertiser obtains statistical data on the features of audiences who have become customers. For example, the percentage of customers categorized into different age groups, such as $\{<18, 19\sim 30, 31\sim40, >40, unknown\}$. With this data, the advertiser can compute the conditional probability $\mathbb{P}(\cdot | Buy)$, which represents the probability of a customer possessing a specific  feature.

A potential customer $c$ can defined by his/her features:
\[
	c = (c_1,~c_2,\dots,~c_n).
\]
where $c_i\in F_i$ for all $i\in [n]$. Here, $c_i$ represents the value of the $i$-th feature of an audience $c$, which can be any element from the set  $\{t_{i,1}, t_{i,2}, \dots, t_{i,m_i}\}$.. Event $\{c_{i}=t_{i,k}\}$ suggests that the customer is type  $t_{i,k}$ regarding feature $F_i$. The advertiser will decide which features are selected.
If a customer is type $t_{i,k}$, then 
\begin{equation}\label{eq:pik}
	P_{i,k} := \mathbb{P}\lf(c_i \equiv t_{i,k} ~|~ Buy \rg)
\end{equation}
 is known to advertiser. We use $P_{i,k}$ to represent the probability of a customer's $i$-th feature being of type $k$.
Likewise, for all $i\in [n]$, the advertiser can observe the probabilities $\{P_{i,1},~P_{i,2},~,\dots, P_{i,m_i}\}$ pertaining to feature $F_i$.

In addition, to facilitate the decision-making process, the platform also provides the statistical data of all audiences. For any  given feature $i$, the advertiser knows the percentage of audiences (regardless of whether they  purchase the product or not) belonging to type $t_{i,k}$ for all $k \in  [m_i]$. So that 
\begin{equation}\label{eq:qik}
	Q_{i,k}: =\mathbb{P}\lf(c_i \equiv t_{i,k}\rg),
\end{equation}
is also known to advertiser. Again, we use $Q_{i,k}$ to represent the probability of an audience's $i$-th feature being of type $k$. Similarly, for all $i\in [n]$, the data $\{Q_{i,1},~Q_{i,2},~,\dots, Q_{i,m_i}\}$ of the $i$-th feature can be observed by the advertiser.

Although many advertising management platforms provide advertisers with data on both $P_{i,k}$ and $Q_{i,k}$, some platforms only offer $P_{i,k}$. To highlight the significance of having access to both $P_{i,k}$ and $Q_{i,k}$ for making informed decisions, let's consider an example where an advertiser is promoting a men's skin care product on a beauty shopping platform. Upon analyzing the data, the advertiser observes that 60\% of the buyers of the skin care product are women, while only 40\% are men. Additionally, there are more active female users (80\%) than male users (20\%) on the shopping platform. Without access to comprehensive statistical information on all audiences on the platform, the advertiser may encounter challenges in formulating an appropriate and effective advertising strategy. Thus, the availability of both $P_{i,k}$ and $Q_{i,k}$ is crucial for advertisers to make informed decisions, as it provides valuable insights into the demographics and preferences of their audiences. 

  
\textbf{The model.}  The strategy $S$ can be represented by the set of selected features:
\[
	{\cal S}:= \{F_{1}^s,~F_{2}^s,~,\dots,~F_{n}^s\},
\]
where $ F_{i}^s\subseteq F_i$ . 
The advertiser can choose multiple possible types from a single feature, meaning that $|F_{i}^s|$ can be greater than $1$. It implies that  the advertiser must select at least one type $t_{i,k}$ for each feature $F_i$. If $F_{i}^s= \emptyset$ for some $i$, it implies that $\mathbb{P}(c\in {\cal S} ) = 0$, meaning the strategy excludes all audiences.  On the contrary, if $F_{i}^s = F_i$, the advertiser accepts any potential customer regardless of their types of feature $F_i$. The advertiser opt to do this if they do not consider the $i$-th feature when designing the advertisement strategy. We refer to the $i$-th feature as ``active''  in an strategy if $F_{i}^s\neq F_i$, and as ``inactive'' otherwise.

The advertiser aims to maximize the sales of their product by targeting audiences with the highest probability of being converted to customers. Suppose that the number of audiences is $N$, given a advertising strategy ${\cal S}$, the sales $q$ can be empirically estimated by  
\begin{equation}\label{eq:expq}
 \E[q] = N \mathbb{P}\lf( Buy\cap c\in {\cal S}\rg) = N \mathbb{P}\lf( Buy ~|~ c\in {\cal S}\rg)  \mathbb{P}\lf(c\in {\cal S}\rg). 
\end{equation}

If the maximum campaign expenditure budget $e$ is predetermined, then value of $\mathbb{P}\lf(c\in {\cal S}\rg)$ is upper bounded. In this case, the problem can be equivalently reformulated as optimizing the advertisement strategy to target  a certain range of audiences while maximizing the estimated conversion rate.  In other words, the advertiser aims to reach a specific set of potential customers within the given budget, while also maximizes the likelihood of successful conversions.
Henceforth, we propose the following model:
\begin{align}\label{eq:opt}
	\underset{{\cal S}}{\max}~ & ~~\mathcal{P}({\cal S}) = \mathbb{P}\lf( Buy ~|~ c\in {\cal S}\rg) \\
	{\rm s.t.} ~~& \qquad \mathbb{P}\lf(c\in {\cal S}\rg) \ge L.\nonumber
\end{align} 
where $L$ is the minimum required range of audiences belonging to the selected set of features. The parameter $L$ is closely related to the advertising campaign expenditure budget $e$. A higher budget $e$ implies larger value of $L$, allowing the advertiser to select a strategy ${\cal S}$ covering a wider range of audiences. 
 
It is important to highlight that a larger number of active features in the advertising strategy or a smaller number of types $t_{i,k}$ included in any features will result in a smaller value of $|{\cal S}|:= \sum_{i=1}^n |F_{i}^s|$, which is the total amount of selected types. Consequently, the value of $\mathbb{P}\lf( Buy | c\in {\cal S}\rg)$ will be higher, while the value of $\mathbb{P}\lf(c\in{\cal S}\rg)$ will be lower. Therefore, the advertiser faces the challenge of striking a delicate balance between maximizing the probability of conversions and achieving a broader coverage of audiences.
It is also important to emphasize  that our proposed advertisement model differs from models that solely focus on optimizing or predicting the probability of click-through rates  (CTR) \citep{shan2016predicting, lee2018estimating, miralles2018novel}. Such models concentrate on maximizing the likelihood of a customer clicking on an ad, given their specific features. On the other hand, our model exhibits similarities with investment models that aim to optimize the return on investment (ROI).  

\section{Model reformulation}\label{sec:remod}

In order to obtain a solvable reformulation of \eqref{eq:opt}, we introduce the following two assumptions:
\begin{assumption}\label{ass:1}
	The events ($c_{i_1}\in F_{i_1}^s$) and ($c_{i_2}\in F_{i_2}^s$) are independent for any ${i_1}\neq {i_2}$, and ${i_1},~{i_2}\in [n]$. 
\end{assumption}

\begin{assumption}\label{ass:2}
	The conditional events ($c_{i_1}\in F_{i_1}^s~| ~Buy $) and ($c_{i_2}\in F_{i_2}^s~| ~Buy$) are independent for any ${i_1}\neq {i_2}$, and ${i_1},~{i_1}\in [n]$. 
\end{assumption}

\begin{remark}
Assumption~\ref{ass:1} is generally reasonable for most demographic features available as target options on a platform since they tend to be statistically independent.  However, it is essential to acknowledge that some features may exhibit high correlation. For instance, features like \texttt{purchase power} and \texttt{monthly shopping frequency} could be strongly related. In such cases, the advertiser must carefully select only one of these correlated features to avoid violating the independence assumptions during the decision-making process. Section~\ref{sec:exp} presents a real-world data example to further illustrate the practical implications of these assumptions in our study.

The validity of Assumption~\ref{ass:2} is heavily influenced by the characteristics of the product being promoted. Some products, like wrinkle creams, may display strong gender and age preferences among potential customers. As a result, there are certain scenarios where Assumption~\ref{ass:2} may not hold, potentially limiting the applicability of our proposed method in such specific cases. It is crucial for advertisers to be aware of these product-specific considerations when utilizing our approach for optimizing their advertising strategies.
\end{remark}

We provide the following theorem for the reformulation of model \eqref{eq:opt}.
\begin{theorem}\label{thm:rfm}
	Suppose that Assumptions~\ref{ass:1}-\ref{ass:2} hold, then problem \eqref{eq:opt} can be reformulated as
	\begin{align}\label{eq:opt2}
	\underset{{\cal S} }{\max}~ & ~~\mathcal{P}({\cal S}) = \frac{\prod_{i=1}^n \PP(c_i\in F_{i}^s~|~ Buy)\cdot \PP( Buy)}{\prod_{i=1}^n\PP(c_i\in F_{i}^s)} \\
	{\rm s.t.} ~~&  \prod_{i=1}^n\PP(c_i\in F_{i}^s) \ge L,\nonumber\\
	&{\cal S}=\{F_{1}^s,~F_{ 2}^s,~\dots,~F_{ n}^s\}.\nonumber
	\end{align}
\end{theorem}
\begin{proof}

We recall the Bayes' theorem to obtain:
\begin{equation}\label{eq:bayes}
	  \PP( Buy ~|~c\in {\cal S}) 
	= \frac{\PP(c\in {\cal S}~|~ Buy)\cdot \PP( Buy)}{\PP(c\in {\cal S})}.
\end{equation}

Since Assumption ~\ref{ass:1} holds, the  denominator of equation \eqref{eq:bayes} and the constraint in problem \eqref{eq:opt} equals to
\begin{equation}\label{eq:deno}
	\PP(c\in {\cal S}) = \prod_{i=1}^n\PP(c_i\in F_{i}^s).
\end{equation}

By Assumption ~\ref{ass:2}, the numerator of equation \eqref{eq:bayes} can be written as
\begin{equation}\label{eq:num}
	\PP(c\in {\cal S}~|~ Buy)\cdot \PP( Buy) = \prod_{i=1}^n \PP(c_i\in F_{i}^s~|~ Buy)\cdot \PP( Buy).
\end{equation}

 Plugging equations \eqref{eq:deno} and \eqref{eq:num} into equation \eqref{eq:bayes} , the right hand side of equation \eqref{eq:bayes} becomes
\[
	\frac{\PP(c\in {\cal S}~|~ Buy)\cdot \PP( Buy)}{\PP(c\in {\cal S})} = \frac{\prod_{i=1}^n \PP(c_i\in F_{i}^s~|~ Buy)\cdot \PP( Buy)}{\prod_{i=1}^n\PP(c_i\in F_{i}^s)} .
\]

\end{proof}

To arrive at the final advertising strategy, the advertiser must make decisions for each $F_{i}^s$ for all $i\in[n]$. To facilitate this process, we present the following proposition.

\begin{proposition}\label{prop:irr}
	Suppose Assumptions~\ref{ass:1}-\ref{ass:2} hold, then we have the following conclusions:
	\begin{enumerate}[1]
		
	\item Denote by ${\cal S}_{i}^s := \{F_1,~F_2,~\dots,~F_{i-1},~F_{i}^s,~ F_{i+1},~\dots, ~F_n\}$ the $i$-th sub-strategy. We have 
	\[
		{\cal P}({\cal S}) = \frac{\prod_{i=1}^{n}{\cal P}({\cal S}_{i}^s)}{\PP(Buy)^{n-1}}.
	\]
	\item Denote by ${\cal S}_{\#i}^s := \{F_{1}^s,~F_{2}^s,~\dots,~F_{i-1}^s,~F_i,~ F_{i+1}^s,~\dots, ~F_{n}^s\}$ the complement of the $i$-th sub-strategy ${\cal S}_{i}^s$. We have  
	\[
		{\cal P}({\cal S}_{\#i}^s )= \frac{\prod_{j=1,j\neq i}^n \PP(c_j\in F_{j}^s~|~ Buy)\cdot \PP( Buy)}{\prod_{j=1,j\neq i}^n\PP(c_j\in F_{j}^s)},
	\]
	and
	\[
		\PP(c\in{\cal S}_{\#i}^s ) = \prod_{i=1,i\neq j}^n\PP(c_i\in F_{i}^s).
	\]	
	\item Suppose ${\cal S}^*:=\{F_{1}^{s*},~F_{2}^{s*},~\dots,~F_{n}^{s*}\}$ is the optimal strategy. Denote by ${\cal I}$ the index set with ${\cal I} = \{i~|~F_{i}^{s*} = F_i,~i\in [n]\}$, and by $\bar{\cal I}$  = $ [n]\setminus {\cal I}$. We have 
	\[
		{\cal P}({\cal S}^*)  = \frac{\prod_{i\in \bar{\cal I}}{\cal P}({\cal S}_{i}^{s*})}{\PP(Buy)^{|{\cal I}|-1}} = \frac{\prod_{i\in \bar{\cal I}} \PP(c_i\in F_{i}^{s*}~|~ Buy)\cdot \PP( Buy)}{\prod_{i\in \bar{\cal I}} \PP(c_i\in F_{i}^s)}.
	\]
	\end{enumerate}
\end{proposition}
\begin{proof}
	We will only prove item 1, because items 2-3 are obvious if item 1 is given. Since
	\begin{align}\label{eq:pr1}
		{\cal P}({\cal S}_{i}^s) = & \frac{\PP(c\in {\cal S}_{i}^s~|~ Buy)\cdot \PP( Buy)}{\PP(c\in {\cal S}_{i}^s)} \nonumber \\
		= & \frac{ \PP(c_i\in F_{i}^s~|~ Buy) \cdot\prod_{j=1,j\neq i}^n \PP(c_j\in F_j~|~ Buy)\cdot \PP( Buy)}{P(c_i\in F_{i}^s )\prod_{j=1,j\neq i}^n\PP(c_j\in F_j )} \nonumber \\
		=& \frac{ \PP(c_i\in F_{i}^s~|~ Buy)\cdot \PP( Buy)}{\PP(c_i\in F_{i}^s)},
	\end{align}
	we have
	\begin{align*}
		\prod_{i=1}^{n}{\cal P}({\cal S}_{i}^s) = & \frac{\PP( Buy)^n \prod_{i=1}^n\PP(c_i\in F_{i}^s~|~ Buy) }{\prod_{i=1}^n\PP(c_i\in F_{i}^s )} \\
		= & \PP( Buy)^{n-1} {\cal P}({\cal S}).
	\end{align*}
	Then the proof of item 1 is concluded.
\end{proof}

Proposition \ref{prop:irr} indicates that the optimization problem \eqref{eq:opt2} is separable, allowing  one to determine the optimal strategy ${\cal S}^*$ by obtaining a series of optimal sub-strategies. Based on Proposition \ref{prop:irr} and equation \eqref{eq:pr1}, we write the $i$-th subproblem as follows:
\begin{align}\label{eq:opt2_2}
	\underset{{\cal S}_{i}^s}{\max}~ & ~~\mathcal{P}({\cal S}_{i}^s) = \frac{ \PP(c_i\in F_{i}^s~|~ Buy)\cdot \PP( Buy)}{\PP(c_i\in F_{i}^s)} \\
	{\rm s.t.} ~~& \qquad \PP(c_i\in F_{i}^s) \ge L_i.\nonumber
\end{align}
where $L_i$ is the required minimum probability for the $i$-th subproblem. Unlike the value of $L$ in problem \eqref{eq:opt}, which can be derived from the advertising campaign expenditure budget $e$, we cannot directly obtain the values of $L_i$ for individual features. While obtaining the values of $L_i$ may be challenging, it is not essential for the optimization process. 

For the purpose of solving the subproblem \eqref{eq:opt2_2}, we introduce the following assumption:
\begin{assumption}\label{ass:ind}
	The events $\{(c_i= t_{i,k})\}_{k=1}^{m_i}$ for any $i\in [n]$ are mutually exclusive and collectively exhaustive (MECE).
\end{assumption}

Based on Assumption \ref{ass:ind}, we have the following two lemmas. The proves of the lemmas are omitted as they are trivial. 

\begin{lemma}
	Suppose Assumption~\ref{ass:ind} holds, then the conditional events $\{(c_i= t_{i,k}|~ Buy)\}_{k=1}^{m_i}$ for any $i\in [n]$ are MECE. 
\end{lemma}

\begin{lemma}
	For any $i\in [n]$, arbitrarily take any ${k_1},~{k_2}\in  [m_i]$, we have \[
	\PP(c_i\in\{t_{i,{k_1}},t_{i,{k_2}}\}) = \PP(c_i=t_{i,{k_1}}) + \PP(c_i=t_{i,{k_2}}),
\]
and  
\[
	\PP(c_i\in\{t_{i,{k_1}},t_{i,{k_2}}\}|~ Buy) = \PP(c_i=t_{i,{k_1}}|~ Buy) + \PP(c_i=t_{i,{k_2}}|~ Buy).
\]
\end{lemma}

As a result, a good procedure for us to set up the solution to a subproblem is to  fill the capability $L_i$ by the types with profit-to-weight ratios $\PP(c_i=t_{i,k}|Buy)/\PP(c_i=t_{i,l})$ from high to low. Based on this idea, we propose Algorithm \ref{alg:sub} for solving the subproblem \eqref{eq:opt2_2}. Next, we will prove that the solution obtained by Algorithm \ref{alg:sub} is optimal.

\renewcommand{\thealgorithm}{\arabic{algorithm}}
\begin{algorithm}[H]
\setcounter{algorithm}{0}
\caption{(Greedy algorithm for  subproblem \eqref{eq:opt2_2})}\label{alg:sub}
	\begin{algorithmic}[1]
	\State Initialization: sort the types $\{t_{i,1},~t_{i,2},~\dots,~t_{i,m_i}\}$ of the $i$-th subproblem by non-increasing ratio $\PP(c_i=t_{i,k}|~Buy)/\PP(c_i=t_{i,k})$ for all $k\in [m_i]$. Set $F_{i}^s = \emptyset$ and ${\cal F}_{i}=\emptyset$. 

	\For{$k=1,2,3,\ldots,m_i$}
		\State Set $F_{i,k}^s = \{t_{i,1},\dots, t_{i,k}\}$ and include $F_{i,k}^s\in {\cal F}_{i}$.
		\If{$\PP(c_i\in F_{i,k}^s)\ge L_i$},
		 break.
		\EndIf
	\EndFor 
	\State Output: $F_{i}^{s*} := F_{i,k}^{s}$ and ${\cal F}_{i}$.
	\end{algorithmic}
\end{algorithm}

\begin{lemma}
	Any strategy $F_{i}^{s*}$ obtained by Algorithm \ref{alg:sub} satisfies that 
	\[
		\frac{\PP(c_i \in F_{i}^{s*}~|~Buy)}{\PP(c_i \in F_{i}^{s*})}\ge 1.
	\]
\end{lemma}
\begin{proof}
If $m_i = 1$, we immediately obtain that $F_{i}^{s*} = \{t_{i,1}\}$ and $\frac{\PP(c_i \in F_{i}^{s*}~|~Buy)}{\PP(c_i \in F_{i}^{s*})} = 1$. 
If $m_i>1$, then there exists at least one $t_{i,k}\in F_i $ such that $\PP(c_i = t_{i,k}~|~Buy)/\PP(c_i = t_{i,k})> 1$. 
Suppose by contradiction that for all $t_{i,k}\in F_i $, we have $\PP(c_i = t_{i,k}~|~Buy)/\PP(c_i = t_{i,k})<1$. Then we obtain that $\frac{\sum_{k=1}^{m_i}\PP(c_i = t_{i,k}~|~Buy)}{\sum_{k=1}^{m_i}\PP(c_i = t_{i,k})}<1$, it contradicts with the fact that $\sum_{k=1}^{m_i}\PP(c_i = t_{i,k}~|~Buy) = 1$ and $\sum_{k=1}^{m_i}\PP(c_i = t_{i,k}) = 1$. Suppose that for all $i\in\{1,~2,~\dots,~n\}$ we have $m_i>1$. 
Hence we conclude that $\PP(c_i = t_{i,1}~|~Buy)/\PP(c_i = t_{i,1})> 1$ for all $i$. 

Suppose that there exists a set $F_{i}^{s*}$ obtained by Algorithm \ref{alg:sub} such that $P(c_i \in F_{i}^{s*}~|~Buy)/P(c_i \in F_{i}^{s*})<1$, and suppose that Algorithm \ref{alg:sub} terminates when $k = K<m_i$. We have 
\[
	\frac{\sum_{k=K+1}^{m_i}\PP(c_i = t_{i,k}~|~Buy)}{\sum_{k=K+1}^{m_i}\PP(c_i = t_{i,k})}
	<\frac{\sum_{k=1}^{K}\PP(c_i = t_{i,k}~|~Buy)}{\sum_{k=1}^{K}\PP(c_i = t_{i,k})}
	=\frac{\PP(c_i \in F_{i}^{s*}~|~Buy)}{\PP(c_i \in F_{i}^{s*})} 
	<1. 
\]

Since $\frac{\sum_{k=1}^{m_i}\PP(c_i = t_{i,k}~|~Buy)}{\sum_{k=1}^{m_i}\PP(c_i = t_{i,k})}  = \frac{1}{1}$, it immediately derives a contradiction that 
\[
	\frac{\sum_{k=K+1}^{m_i}\PP(c_i = t_{i,k}~|~Buy)}{\sum_{k=K+1}^{m_i}\PP(c_i = t_{i,k})} 
	< \frac{\sum_{k=1}^{m_i}\PP(c_i = t_{i,k}~|~Buy)}{\sum_{k=1}^{m_i}\PP(c_i = t_{i,k})} 
	<\frac{\sum_{k=1}^{K}\PP(c_i = t_{i,k}~|~Buy)}{\sum_{k=1}^{K}\PP(c_i = t_{i,k})}.
\]

Therefore, we have $\frac{\PP(c_i \in F_{i}^{s*}~|~Buy)}{\PP(c_i \in F_{i}^{s*})} >1 $ for all $i$ with $m_i>1$. 
\end{proof}

\begin{theorem}\label{thm:subopt}
	Suppose Assumption \ref{ass:ind} holds, then the solution $F_{i}^{s*}$ obtained by Algorithm \ref{alg:sub} is a global optimal solution to subproblem \eqref{eq:opt2_2}.
\end{theorem}
\begin{proof}
	
	Since any set $F_{i}^{s*}$ obtained by Steps 1$\sim$7 of Algorithm \ref{alg:sub} satisfies that $\PP(c_i \in F_{i}^{s*}~|~Buy)/\PP(c_i \in F_{i}^{s*})\ge 1$ and $\PP(c_i\in F_{i}^{s*})>L_i$. 
	By Assumption \ref{ass:ind}, we have that for all $j \in [m_i]$, $t_{i,k}$ are mutually exclusive, then 
	\[
		\mathcal{\PP}({\cal S}_{i}^s) = \frac{\sum_{t_{i,k}\in F_{i}^{s*}}\PP(c_i = t_{i,k}~|~Buy)}{\sum_{t_{i,k}\in F_{i}^{s*}}\PP(c_i = t_{i,k})}\cdot \PP(Buy).
	\]
	
	Since the types are ordered by non-increasing ratio $\PP(c_i= t_{i,k}~|~Buy)/\PP(c_i= t_{i,k})$, then, started from an empty set, the sub-strategy ${\cal S}_{i}^s$ is item-by-item filled by Algorithm \ref{alg:sub} until the first item that satisfies $\PP(c_i \in F_{i}^s)\ge L_i$. For the sake of brevity, we use the notation $Q(x) = \sum_{k=1}^{x} \PP(c_i = t_{i,k})$. 
We denote $\beta$ the indicator corresponding to this break item:
\[
	\beta = \underset{x}{\arg\min}\{Q(x) > L_i\}.
\]
Based on the fact that if $a,~b,~c,~d>0$ and $\frac{a}{b}\ge\frac{c}{d}$, then $\frac{a}{b}\ge\frac{a+c}{b+d}\ge\frac{c}{d}$, one can easily obtain that the sequence of the objective values $\mathcal{P}({\cal S}_{i}^s) $ generated by Algorithm \ref{alg:sub} is non-increasing. In consequence, the algorithm terminates once the constraint in  subproblem  \eqref{eq:opt2_2} holds.  Since the ratio of $\PP(c_i = t_{i,1}~|~Buy)/\PP(c_i = t_{i,1})$ is the largest, the largest objective value of $\mathcal{P}({\cal S}_{i}^s) $ is reached if the strategy uniquely include the first item.  	

	Suppose by contradiction that $F_{i}^{s*}$ obtained by Algorithm \ref{alg:sub} is not optimal, then there exist $t_{i,{k_1}}$ and $t_{i,{k_2}}$ such that ${k_1}>\beta$, ${k_2}<\beta$, we have $\PP(c_i=t_{i,{k_1}})> \PP(c_i = t_{i,{k_2}})$ and $\mathcal{P}\lf((F_{i}^{s*}\setminus\{t_{i,{k_2}}\})\cup\{t_{i,{k_1}}\}\rg) > \mathcal{P}\lf(F_{i}^{s*}\rg)$. However, based on the fact that if $a,~b,~c,~d>0$ and $\frac{a}{b}>\frac{c}{d}$, then $\frac{a}{b}>\frac{a+c}{b+d}>\frac{c}{d}$, we derive a contradiction that $\PP(c_i=t_{i,{k_1}})<\PP(c_i = t_{i,{k_2}})$.

	Hence, we can conclude that the solution generated by Algorithm \ref{alg:sub} is globally optimal to subproblem  \eqref{eq:opt2_2}. 
	
\end{proof}

We use the following example to illustrate how Algorithm \ref{alg:sub} solves subproblem \eqref{eq:opt2_2}. 
\begin{example}
	Consider the case $F_i = \{t_{i,1}, t_{i,2}, t_{i,3}, t_{i,4}, t_{i,5}, t_{i,6}\}$ with the data presented in the Table \ref{table1}. Suppose $\PP(Buy) = B$.

\begin{table}
	\renewcommand\arraystretch{1.3}
	\setlength{\abovecaptionskip}{2pt}
	\setlength{\belowcaptionskip}{0pt}
	\centering
	\caption{Numerical example of finding the solution to the subproblem}
	\label{table1}
	\scriptsize
	\begin{tabular*}{\textwidth}{@{\extracolsep{\fill}}lccccl@{\extracolsep{\fill}}}
	\hline
	 & $t_{i,k}$ & $P(c_i=t_{i,k})$ & $P(c_i=t_{i,k}|Buy)$ & $P(c_i=t_{i,k}|Buy)/P(c_i=t_{i,k})$ &  \\
	\hline
	 & $t_{i,1}$ & ~7.28\% & 16.27\% & 2.20  & \\
	 & $t_{i,2}$ & 26.00\% & 49.92\% & 1.92  & \\
	 & $t_{i,3}$ & 27.75\% & 19.88\% & 0.72  & \\
	 & $t_{i,4}$ & 19.10\% & ~7.63\% & 0.39  & \\
	 & $t_{i,5}$ & 12.50\% & ~2.76\% & 0.22  & \\
	 & $t_{i,6}$ & ~7.27\% & ~3.64\% & 0.50  & \\
	\hline
	\end{tabular*}
\end{table}
	
According to the last column of the table, we sort ${t_{i,1}, t_{i,2}, t_{i,3}, t_{i,6}, t_{i,4}, t_{i,5}}$ by the ratio $\PP(c_i=t_{i,k}|Buy)/\PP(c_i=t_{i,k})$. We include $t_{i,1}$ in the strategy set. If $L_i=0$, Algorithm \ref{alg:sub} terminates when $k=1$, giving the optimal strategy ${\cal S}_{i}^{s*} = {t_{i,1}}$. The value of the objective function $\mathcal{P}({\cal S}_{i}^{s*})$ is $2.20B$. If $L_i = 30\%$, Algorithm \ref{alg:sub} terminates when $k=2$. We get the optimal strategy ${\cal S}_{i}^{s*} = {t_{i,1}, t_{i,2}}$, and the value of the objective function $\mathcal{P}({\cal S}_{i}^{s*})$ is $(16.27\% +49.92\%)\cdot B/(7.28\% + 26.00\%) = 1.99B$.
\end{example}

The time complexity of sorting algorithm is $O(m_i log(m_i))$, and the time complexity of the loop in Algorithm \ref{alg:sub} is $O(m_i)$. Hence, the total time complexity of Algorithm \ref{alg:sub} is $O(m_i log(m_i))$. 
An immediate difficulty lies in the fact that the parameter $L_i$ cannot be decided locally. 
Hence, we set $L_i=1$ and record all the possible optimal solutions in $S_i$, and the total time complexity is $O(\sum_{i=1}^n m_i log(m_i))$. Henceforth, problem \eqref{eq:opt2} can be rewritten as:
\begin{align}\label{eq:opt3}
	\underset{\{F_{1}^s,~F_{2}^s,~\dots,~F_{n}^s\}}{\max}~ & ~~ \frac{\prod_{i=1}^n \PP(c_i\in F_{i}^s~|~ Buy)\cdot \PP( Buy)}{\prod_{i=1}^n\PP(c_i\in F_{i}^s)}\nonumber  \\
	{\rm s.t.} ~~&  \prod_{i=1}^n\PP(c_i\in F_{i}^s) \ge L,\\
	& F_{i}^s = F_{i,k}^s \in {\cal F}_i, \quad i  \in[n],\quad k\in [m_i]. \nonumber
\end{align}

Since all the probabilities appeared in problem \eqref{eq:opt3} are nonnegative, we can take the logarithm of both the objective function and the constraint. Once $F_{i}^s$ for all $i\in\{1,2,\dots,~n\}$ is obtained, we can further convert problem \eqref{eq:opt3} to:
\begin{align}\label{eq:optks}
	\underset{x}{\max} ~ & ~~\Phi(x) :=\sum_{i=1}^n \sum_{k=1}^{m_i}(p_{ik}-q_{ik}) x_{ik} + \beta \nonumber  \\
	{\rm s.t.} ~~& \qquad \sum_{i=1}^n\sum_{k=1}^{m_i} -q_{ik} x_{ik} \le -l, \\
			 	 & \qquad \sum_{k=1}^{m_i}x_{ik}\le 1, \quad i  \in[n] ,\nonumber \\
				 & \qquad  x_{ik}\in\{0,1\}, \quad  i \in[n],\quad k\in [m_i].\nonumber 
\end{align}
where $p_{ik} = \log(\PP(c_i\in F_{i,k}^{s}~|~ Buy))$, $q_{ik} = \log(\PP(c_i\in F_{i,k}^{s}))$, $\beta = \log(\PP( Buy))$, and $l = \log(L)$. Problem \eqref{eq:optks} is a multiple choice knapsack problem (MCKP). Given a set of items, each with a weight and a value,  MCKP aims to decide for each item to be included or not so that the total value is maximized, and the total weight is no more than  a given limit. In our case, the values of items are $\{p_{ik} - q_{ik}\}$, the weights of items are $\{-q_{ik}\}\ge 0$, and the weight limit is $-l\ge 0$. 

\subsection{Problem \eqref{eq:optks} simplification}
We now give some assumptions which do not lose generality but simplify problem \eqref{eq:optks}. 
\begin{assumption}\label{ass:simp}
	The parameters in problem \eqref{eq:optks} satisfy:
	\begin{enumerate}
	\item $p_{ik}-q_{ik}\ge 0$ for any $i\in [n]$ and $k\in [m_i]$;
	
	\item For any $i\in [n]$, we have
	\begin{equation}\label{eq:ass_simp1}
	\begin{aligned}
		p_{i1}-q_{i1} & \le p_{i2}-q_{i2} & \le \dots & \le p_{im_i}-q_{im_i}, \\
		-q_{i1} & \le -q_{i2} & \le \dots & \le -q_{im_i}.
	\end{aligned}
	\end{equation}

	\item For any $i\in [n]$, we have
	\begin{equation}\label{eq:ass_simp2}
		\frac{p_{im_i}-q_{im_i}}{-q_{im_i-1}}<\frac{p_{im_i-1}-q_{im_i-1}}{-q_{im_i-1}}<\dots <\frac{p_{i1}-q_{i1}}{-q_{i1}}.
	\end{equation}
	\item Based on \eqref{eq:ass_simp1}, we have
	\begin{equation}\label{eq:ass_simp3}
		\sum_{i=1}^n q_{i1} < l.
	\end{equation}
	\end{enumerate}
\end{assumption}

\begin{remark}
	Assumption \ref{ass:simp} does not lose generality. To see this, 
	\begin{enumerate}
	\item we rearrange second equation of  \eqref{eq:ass_simp1} so that $-q_{i1}  \le -q_{i2}  \le \dots  \le -q_{im_i}$ holds. Suppose that $-q_{ik}  \le -q_{ik+1}$ but $p_{ik}-q_{ik}  \ge p_{ik+1}-q_{ik+1} $, and $x_{ik}=0$ and $x_{ik+1}=1$ hold in an optimal solution $x$. If there is a solution $x'$ with $x'_{ik}=1$ and $x'_{ik+1}=0$ also satisfy the constraints in \eqref{eq:optks}, but
	\[
		 p_{ik+1}-q_{ik+1}x'_{ik+1} + p_{ik}-q_{ik}x'_{ik}>p_{ik+1}-q_{ik+1}x_{ik+1} + p_{ik}-q_{ik}x_{ik}.
	\] 
	This contradicts the optimality of $x$. Thus $x_{ik}=0$ can be assumed and $x_{ik}$ can be deleted from \eqref{eq:optks}; 
	\item Let $-q_{ik}  \ge -q_{ik+1}$ and $p_{ik}-q_{ik}  \ge p_{ik+1}-q_{ik+1} $. Assume that 
	\[
		\frac{p_{ik}-q_{ik}}{-q_{ik}}\ge \frac{p_{ik+1}-q_{ik+1}}{-q_{ik+1}},
	\]
	Suppose that $x$ is an optimal solution such that $x_{ik}=0$ and $x_{ik+1}=1$. Consider $x'$ with $x'_{ik}=1$ and $x'_{ik+1}=0$, and it is easy to know that $x'$ is also a feasible solution and $\Phi(x')\ge \Phi(x)$. Thus $x_{ik+1}=0$ can be assumed and  $x_{ik+1}$ can be deleted from \eqref{eq:optks}; 
	\item suppose that \eqref{eq:ass_simp3} does not hold, then it is easy to conclude that $x_{11}=x_{21}=\dots=x_{n1}=1$, and $x_{ik}=0$ for all $k>0$ is trivially optimal. 
	\end{enumerate}
\end{remark}

To close this section, we present the following proposition to characterise the relation between $\Phi(x)$ and $e$.

\begin{proposition}\label{prop:mono}
	The objective function $\phi(e)$ is monotonically decreasing with respect to $e$. 
\end{proposition}

\begin{proof}
	We will use the weak duality theorem to prove this proposition. The linear relaxation of \eqref{eq:optks} is:
\begin{align}\label{eq:optlr}
	\underset{x}{\max} ~ & ~~ \bar{\Phi}(x) + \beta:=\sum_{i=1}^n \sum_{k=1}^{m_i}(p_{ik}-q_{ik}) x_{ik} + \beta \nonumber  \\
	{\rm s.t.} ~~& \qquad \sum_{i=1}^n\sum_{k=1}^{m_i} -q_{ik} x_{ik} \le -l, \\
			 	 & \qquad \sum_{k=1}^{m_i}x_{ik}\le 1, \quad i  \in[n] ,\nonumber \\
				 & \qquad  x_{ik}\ge 0, \quad  i \in[n],\quad k\in [m_i].\nonumber 
\end{align}

The dual problem of \eqref{eq:optlr} is
\begin{align}\label{eq:optd}
	\underset{\mu}{\min} ~ & ~~ \bar{\Phi}^*(\mu ):=\sum_{i=1}^n \mu_i -l\mu_0 \nonumber  \\
	{\rm s.t.} ~~& \qquad \mu_i\ge(p_{ik}-q_{ik}) +q_{ik}\mu_0 , \quad  i \in[n],\quad k\in [m_i], \\
			 	 & \qquad \mu_i\ge 0, \quad i  \in[n] ,\nonumber \\
				 & \qquad  \mu_0\ge 0.\nonumber 
\end{align}
	where $\mu = (\mu_0,\mu_1, \dots, \mu_n)\in \R^{n+1}$. It is obvious that $\bar{\Phi}^*(\mu^*)$ is monotonically decreasing with $l$, given an optimal value $\mu^*$. Based on the weak duality, we have the following relation between the optimal objective values of problems \eqref{eq:optd},  \eqref{eq:optlr}, and \eqref{eq:optks}:
	\[
		\bar{\Phi}^*(\mu^*)\ge \bar{\Phi}^*(x^*)\ge \Phi(x^*).
	\]

Concerning the connections between $\phi(x)$ and $\Phi(x)$, along with the relationships between $e$ and $l$, we arrive at the conclusion that $\Phi(x^*)$ exhibits a monotonically decreasing behavior with respect to $e$.
	
\end{proof}

\section{Empirical study}\label{sec:exp}

By the virtue of above, we provide the following procedure to generate an advertising strategy with given budget constraint: 

\begin{algorithm}[H]
\setcounter{algorithm}{0}
\caption{(Procedure to generate the optimal advertising strategy)}\label{alg:main}
	\begin{algorithmic}[1]
	\State Inputs: the statistic data $P_{i,k}$ and $Q_{i,k}$ for all $k\in[m_i]$ and $i\in[n]$; advertising expenditure budget $e$.
	\State For all $i\in[n]$, Set $L_i= 1$ and use Algorithm \ref{alg:sub} to obtain ${\cal F}_i$.
	\State Calculate $p_{ik} = \log(\PP(c_i\in F_{i,k}^{s}~|~ Buy))$, $q_{ik} = \log(\PP(c_i\in F_{i,k}^{s}))$, and $l = \log(L)$. Formulate the MCKP \eqref{eq:optks}.
	\State Simplify \eqref{eq:optks} according to Assumption \ref{ass:simp}.
	
	\State Solve \eqref{eq:optks} and obtain optimal solution $x^*$.
	\State For all $i\in[n]$, set $F_{i}^{s*} = F_{i,k}^{s}$ if $x_{ik}^* = 1$.
	\State Output: ${\cal S}^* = \{F_{1}^{s*}, F_{2}^{s*}, \dots, F_{n}^{s*}\}$.
	\end{algorithmic}
\end{algorithm}

There are many algorithms have been proposed for solving the Multiple-Choice Knapsack Problem (MCKP), including the branch-and-bound algorithm \cite{sinha1979multiple, dyer1984branch}, the linear programming method \cite{dyer1984n, zemel1980linear}, and hybrid algorithms \cite{ibaraki1978multiple, bean1990hybrid}. 
Although the MCKP is known to be NP-complete, most of the aforementioned methods can efficiently find an optimal solution within a worst-case computation time of $O(n\log n)$ \cite{nakagawa2001calculating}. In this paper, we opt to utilize the implementation proposed by \cite{ibaraki1978multiple} to solve \eqref{eq:optks}. 
Specifically, we firstly solve the LP relaxation \eqref{eq:optlr} and obtain an approximate solution , which will then be used as an initial point of an exact branch-and-bound algorithm to obtain the optimal solution. 

To conduct our empirical study, we gathered statistical data from Tmall for a fitness nutrition product offered by WildFieldHealth\textsuperscript{\texttrademark}, an Australian fitness nutrition brand. Since 2019, WildFieldHealth\textsuperscript{\texttrademark} has been selling its products to Chinese customers through Tmall.com, a Chinese-language business-to-consumer (B2C) online retail platform operated by Alibaba Group.
The statistical dataset on audiences were derived from historical click-through data, comprising over 50 million samples. 
The statistical dataset on customers is based on $20,708$ samples and comprises twenty-four features categorized into four groups: Consumption Behavior, Interests, Demographic features, and Behavioral preference. These datasets consist of columns representing various features related to customers and their shopping behavior. For a more comprehensive understanding of these features, detailed explanations can be found in Tables \ref{tab:pd1} and \ref{tab:pd2} in the appendix, where the size of each feature is indicated in parentheses next to its name.

As previously discussed in Section \ref{sec:remod}, the optimal advertising strategy is influenced by the expenditure budget $e$. To illustrate the impact of the budget on the advertising strategy, we conducted experiments using $50$ different values of $L$, ranging from $0$ to $1$. Notably, this is the only parameter considered in our experiment. The average computation time per repetition is $12.200321$ seconds.



\begin{figure}[H]
  \centering
  \includegraphics[scale=0.4]{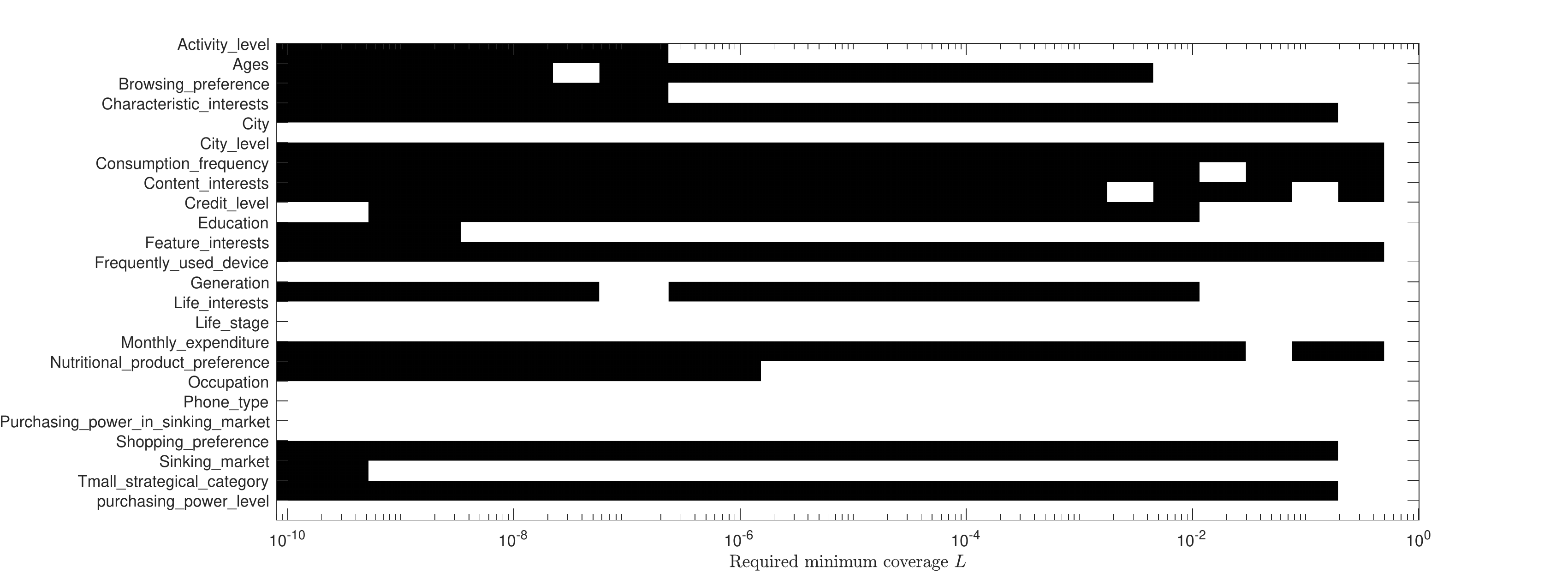}
  \caption{\small The optimal strategies with the correlated features}
  \label{fig:os}
\end{figure}

Figure \ref{fig:os} provides a visual representation of the optimal strategies for different values of $L$, with the selected features highlighted using black squares. Notably, certain features such as \texttt{Ages}, \texttt{Generation}, and \texttt{Life stage} are intuitively highly correlated. Concerning that the coexistence of these features within a strategy may violate Assumption \ref{ass:1}, which asserts feature independence. This violation could potentially lead to inaccuracies in audience targeting. 
Similarly, features such as \texttt{Monthly expenditure}, \texttt{Credit level}, \texttt{Purchasing power in sinking market}, and \texttt{Purchasing power level} may also exhibit interdependencies. Their inclusion in a strategy could result in suboptimal targeting. 

\begin{figure}[H]
  \centering
  \includegraphics[scale=0.4]{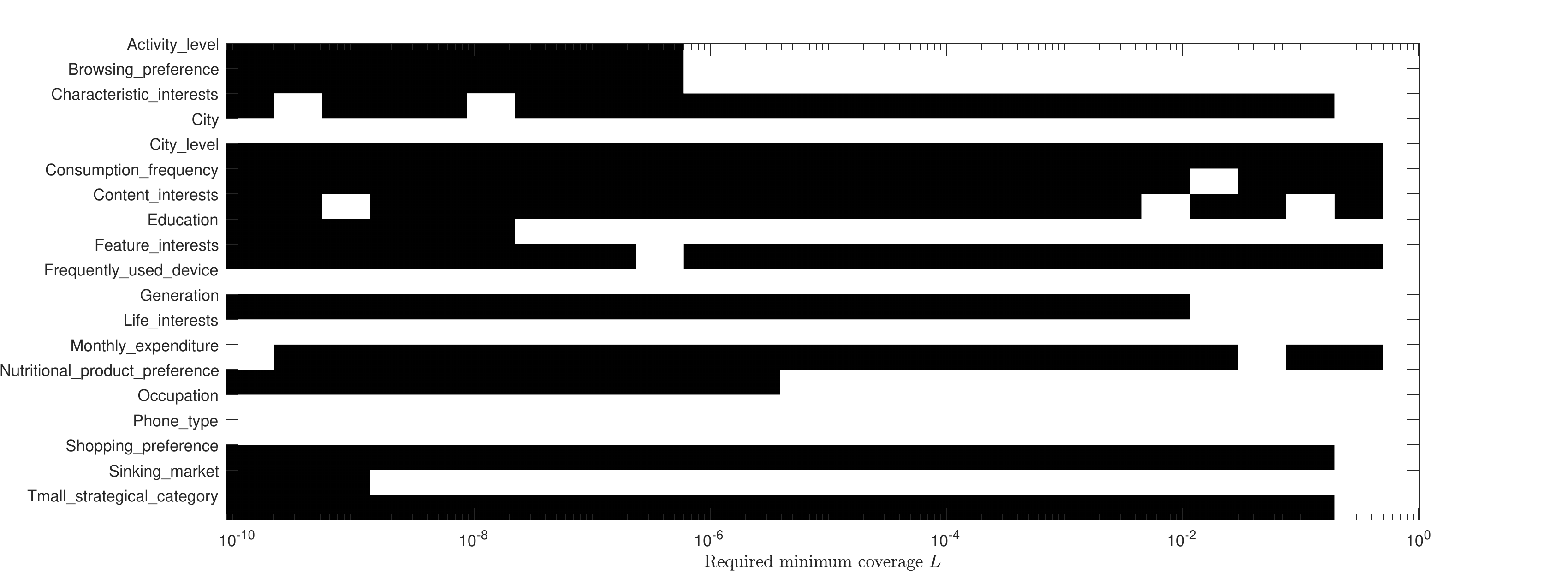}
  \caption{\small The optimal strategies without the correlated features}
  \label{fig:oso}
\end{figure}

We observe that \texttt{Ages} and \texttt{Generation} co-exist in Figure \ref{fig:os}, as well as \texttt{Monthly expenditure} and \texttt{Credit level}.
To ensure Assumption \ref{ass:1}, a strategy refinement process becomes necessary. In this regard, we exclude certain features. Specifically, \texttt{Ages}, \texttt{Life stage}, \texttt{Credit level}, \texttt{Purchasing power in sinking market}, and \texttt{Purchasing power level} are removed. This selection is based on their relatively infrequent appearance in the optimal solutions illustrated in Figure \ref{fig:os}. This strategy adjustment is implemented to maintain the integrity of the assumption of feature independence and thereby enhance the reliability of the model in audience targeting. Again, by using Procedure \ref{alg:main}, we obtain the optimal strategies without the correlated features, which is illustrated in Figure \ref{fig:oso}.

\begin{figure}[H]
  \centering
  \includegraphics[scale=0.4]{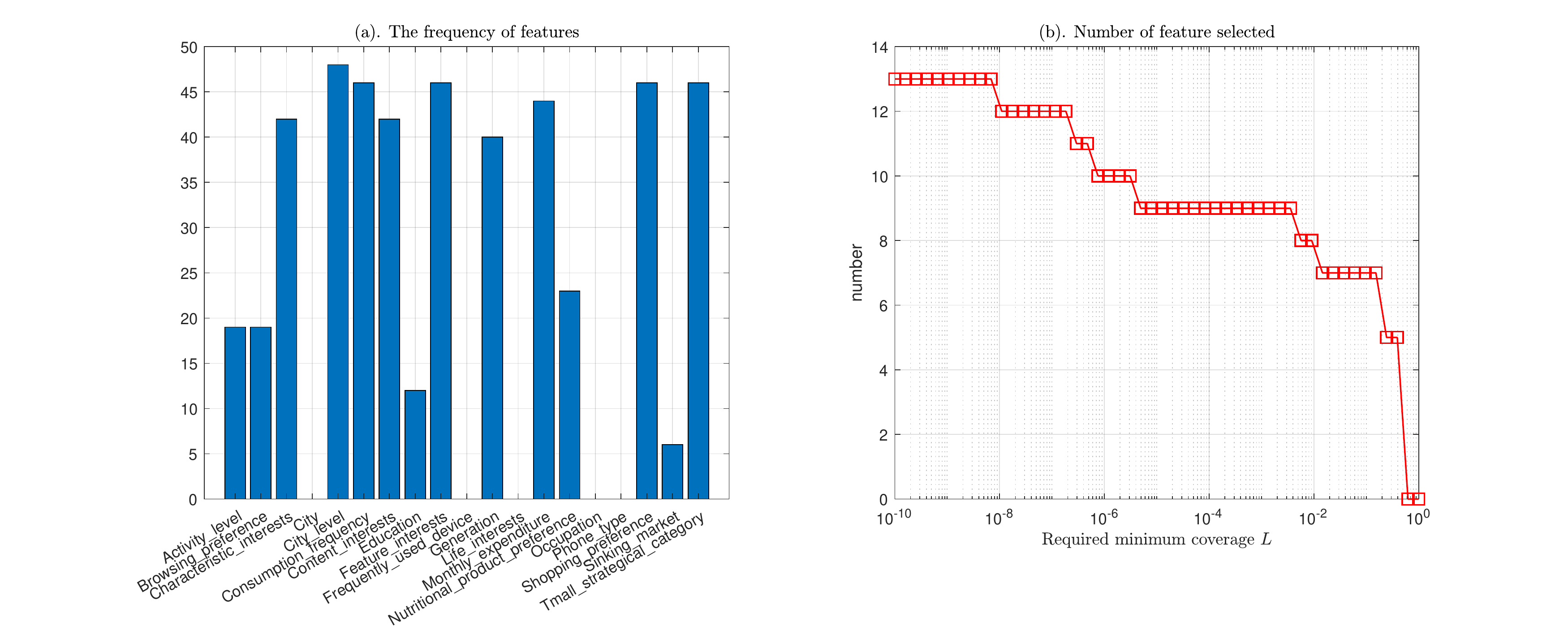}
  \caption{\small Frequency of feature selected}
  \label{fig:fq}
\end{figure}

Figure \ref{fig:fq} (a) presents the frequency distribution of feature selections across all strategies. Notably, the feature \texttt{City level} emerges as the most commonly chosen among all features, followed by \texttt{Consumption frequency}, \texttt{Feature interests}, \texttt{Shopping preference}, and \texttt{Tmall strategical category}. These particular features hold direct relevance to the advertised product and exhibit a strong correlation with customer purchasing behavior for the fitness nutrition product. Consequently, targeting customers based on these high-frequency features presents a promising and optimal approach for the advertiser. This pattern suggests that aligning the advertising strategy with these features is likely to yield favorable results and enhance the overall effectiveness of the campaign.

As the advertising budget $e$ decreases, a corresponding decrease is observed in the value of $L$, as shown in Figure \ref{fig:fq} (b). This results in the optimal advertising strategy encompassing a higher number of active features. Augmenting the advertising budget $e$ leads to a reduction in the number of active features within the optimal strategy. While a larger budget can potentially drive higher sales, the cost-effectiveness may not be guaranteed. Further analysis reveals that the maximum count of activated features in any strategy remains at $13$, out of the total pool of $20$ available features. As $L$ exceeds $1.46\%$, only 7 features maintain their activation status, and when surpassing the $62.51\%$ threshold, no feature remains active. This observation is particularly relevant considering the substantial audience base on Tmall, because targeting more than half of these potential customers might incur excessive costs and inefficiencies. Therefore, the advertiser must carefully select the advertising budget $e$ to maximize profit. 

\begin{figure}[H]
 \makebox[\textwidth][c]{\includegraphics[width=1.0\textwidth]{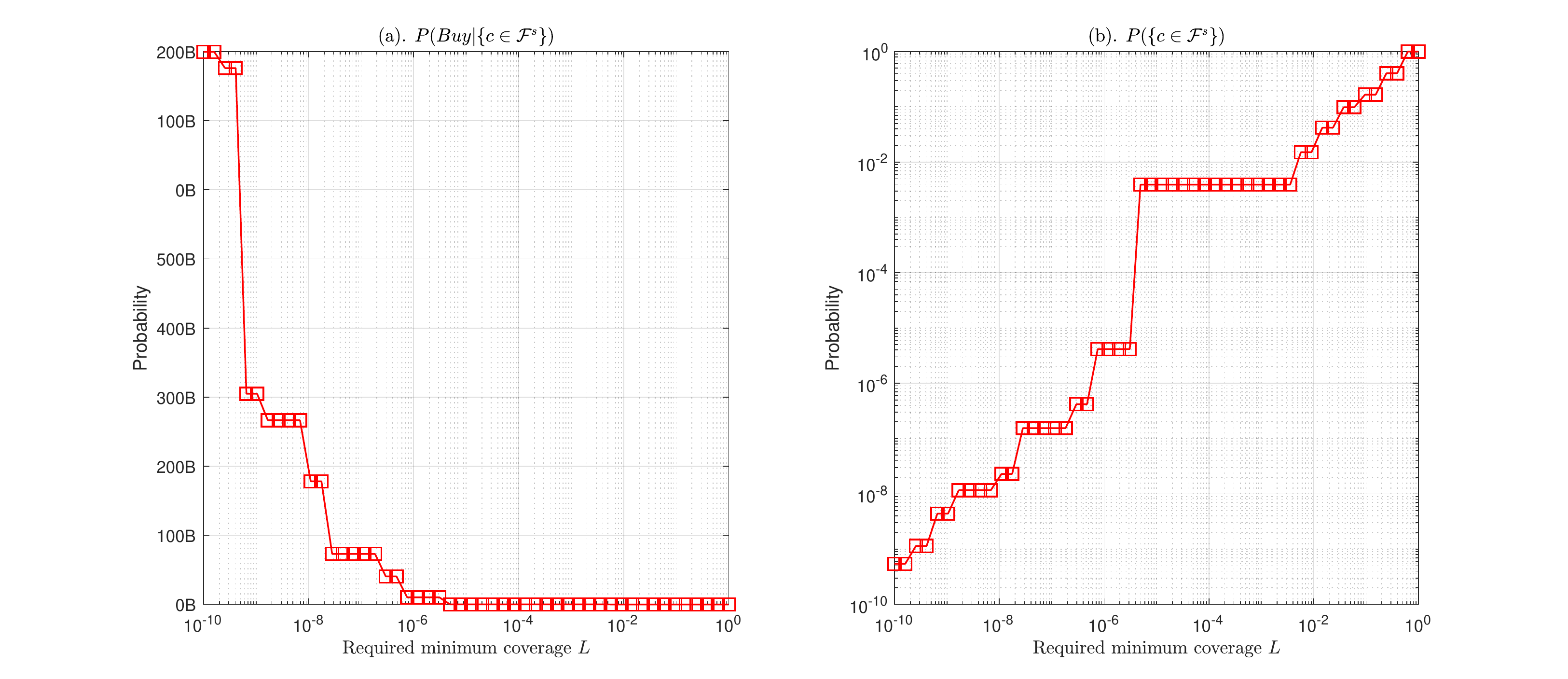}}
  \caption{\small Result of experiments}
  \label{fig:exp}
\end{figure}

Figure \ref{fig:exp} illustrates the results of optimizing problem \eqref{eq:obj0} for different values of $L$. We do not estimate the value of $\PP(Buy)$. Instead, we simply set $\PP(Buy)= B$. It is represented in y-axis of sub-graphs (a). 

As the size of the targeted potential customer group ($\PP({c \in \mathcal{F}_s})$ in sub-graph (b)) decreases, the probability of a potential customer buying the product ($\PP(Buy | {c \in \mathcal{F}^s})$ in sub-graph (a)) increases. However, the probability of selling the product to a potential customer ($\PP(Buy \cap {c \in \mathcal{F}_s})$ as shown in sub-graph (c)) decreases monotonically with respect to $L$. The result aligns with Proposition \ref{prop:mono}.


\section{Conclusions}\label{sec:conc}

In this paper, we develop a modelling framework that classifies audiences based on their distinctive features, which enables us to identify the optimal set of target features that maximizes the advertiser's profit. 
To achieve this objective, we formulate an optimization problem that determines the optimal advertising strategy. To make the problem computationally feasible, we introduce independent assumptions and propose a procedure, which is based on the greedy algorithm,  linear programming, and branch-and-bound algorithm, to fobtain an optimal strategy. We further validate the proposed models by conducting an empirical study using real-world data, which leads to the following key findings:
\begin{enumerate}[1]
\item The empirical study shows that the proposed method effectively optimizes the advertisement strategy under a given expenditure budget. Specifically, a lower budget results in a strategy with more active features, leading to a smaller target group. In contrast, a higher budget typically results in a strategy with fewer active features, leading to a larger target group and higher sales. However, a higher budget may not necessarily lead to a more cost-effective strategy compared to a lower budget. 
\item The \texttt{City level} emerges as the most commonly chosen among all features, followed by \texttt{Consumption frequency}, \texttt{Feature interests}, \texttt{Shopping preference}, and \texttt{Tmall strategical category} are the critical and frequently selected features in optimal strategies as they provide the most information on the probability of customers purchasing the  the nutrition product to be promoted. Moreover, selecting only one feature among highly correlated features is suggested, as selecting more correlated features may lead to less cost-effective strategies.
\end{enumerate}

Admittedly, there are still some limitations in our study. 
The assumptions we make, though aimed at addressing the issue of feature independence, could be still too strong. While we introduce approaches to mitigate the violations, the resulting strategies may inadvertently be suboptimal. Despite our empirical feature elimination efforts to preserve independence assumptions, it remains uncertain whether the interdependencies among the retained features persist. Also, we do not present a method to concurrently optimize both the expenditure budget and advertising strategy.  Moving forward, future studies will aim to address these limitations by considering the intricacies of feature interrelations and jointly optimizing the expenditure budget and advertising strategy.

%
%
%
%
\bibliographystyle{apalike}
\bibliography{bibfile}

\begin{appendices}

\section{Appendix: Feature details}\label{ap:fd}

\begin{landscape}
\begin{table}[]
\caption{Description of the statistical Dataset: Consumption Behavior and Interests}\label{tab:pd1}
\footnotesize
\begin{tabular}{p{0.1\linewidth} | p{0.2\linewidth}| p{0.6\linewidth}}
\toprule
Category & Features & Description \\ \midrule

\multirow{8}{*}{\begin{tabular}[c]{@{}l@{}}Consumption \\ Behavior\end{tabular} }  & Activity level(8) & The count of visit times of customers in the last 30 days. \\ \cmidrule(l){2-3} 
	& Consumption frequency(6)  &The average number of monthly orders placed by customers on Tmall in the last year. \\ \cmidrule(l){2-3} 
	& Credit level(11) & The purchasing credit level of customers when using credit services on Tmall. \\ \cmidrule(l){2-3} 
	& Monthly expenditure(6) &  The average amount spent by customers on Tmall in the last year. \\ \cmidrule(l){2-3} 
    & Purchasing power in sinking market(6) & It measures the purchasing power stratification of sinking markets, based on customer purchasing behavior in the last 6 months. \\ \cmidrule(l){2-3} 
	& Purchasing power level(7) & Customer's purchasing power level is calculated based on their browsing, searching, purchasing, and other behaviors on Tmall. \\ \cmidrule(l){2-3} 
	& Sinking market(6) &  It measures the shopping frequency of customers in tier 4-6 cities and towns. \\ \cmidrule(l){2-3} 
	& Tmall strategical category(9) &  Customers are categorized based on their purchasing preferences. \\ 
\midrule
\multirow{3}{*}{Interests}   & Characteristic interests(7) &  The characteristic interest groups to which the user belongs are obtained comprehensively based on the customer's behavior on Tmall. \\ \cmidrule(l){2-3} 
	& Content interests(6) & Customers' browsing preferences for different forms of content in the last 30 days. \\ \cmidrule(l){2-3} 
	& Feature interests(42) & Tmall categorizes feature interest groups based on the customer's behavior in Tmall. \\ \cmidrule(l){2-3} 
	& Life interests(28) & Tmall.com predicts the life interests   categories according to the behaviors in different scenarios on Hand Taobao,   from the perspectives of style/life stage/purchasing power/interest   preferences. \\ \bottomrule
\end{tabular}
\end{table}
\end{landscape}

\begin{landscape}
\begin{table}[]
\caption{Description of the statistical Dataset: Demographic features and Behavioral preference}\label{tab:pd2}
\footnotesize
\begin{tabular}{p{0.1\linewidth} | p{0.2\linewidth}| p{0.6\linewidth}}
\toprule
Category & Features & Description \\ \midrule
\multirow{8}{*}{\begin{tabular}[c]{@{}l@{}}Demographic \\ features\end{tabular}}  & Ages(7) & The age range of the customer. Tmall obtains this data by analyzing the customer's behavior compared to the entire customer group.  \\ \cmidrule(l){2-3} 
	& City(19) & The city where the customer lives. The data is predicted based on the analysis of the customer's behavior. \\ \cmidrule(l){2-3} 
	& City level(7) &  The city where the shipping address is most frequently used by customers in the last 180 days. \\ \cmidrule(l){2-3} 
	& Education(8) & The education level of the customer. \\ \cmidrule(l){2-3} 
	& Generation(6) & The decade in which the customer was born. Tmall obtains this data by analyzing the customer's behavior compared to the entire customer group. \\ \cmidrule(l){2-3} 
	& Life stage(8) &  The age and stages of the customer currently going through. \\ \cmidrule(l){2-3} 
	& Occupation(9) & The occupation of the customer \\ \cmidrule(l){2-3} 
 	& Phone type(10) &  The phone model type of the customer used to access Tmall. \\ \midrule
\multirow{5}{*}{\begin{tabular}[c]{@{}l@{}}Behavioral \\ preference\end{tabular} } & Browsing preference(8) & The customer's browsing preferences for different forms of content in the last 30 days. \\ \cmidrule(l){2-3} 
	& Frequently used device(4) & Tmall counts the frequency of each device used by the customer within 90 days, and the device with the highest frequency is taken as the frequently used device of a customer.  \\ \cmidrule(l){2-3} 
	& Nutritional product preference(29) & Tmall predicts nutritional product preferences based on the customer's search preferences for nutritional products on Tmall in the last 7 days, click preferences in the last 15 days, favorite preferences in the past 90 days, and purchase preferences in the last 180 days.\\ \cmidrule(l){2-3} 
	& Shopping preference(9) & Based on the customer's interactive behavior on Tmall in the last 180 days and the feature information of the purchased products, Tmall classifies and scores the customers into 5 tiers.\\ \bottomrule
\end{tabular}
\end{table}
\end{landscape}
\end{appendices}

\end{document}